\documentclass[a4paper,11pt]{article}
\pdfoutput=1 

\usepackage{jheppub} 

\usepackage[T1]{fontenc} 

\title{\boldmath Modifications of CMB spectrum by nonextensive statistical mechanics}

\author{Yang Liu}

\affiliation[a]{School of Physics and Astronomy, University of Nottingham, Nottingham NG7 2RD, UK}
\affiliation[b]{Nottingham Centre of Gravity, University of Nottingham, Nottingham NG7 2RD, UK}

\emailAdd{yang.liu@nottingham.ac.uk}

\abstract{Cosmic microwave background radiation can supply us some most significant parts of the information on the universe. Some researchers believe that The gravitational system cannot be decribed by the standard statistical mechanics. In this article we apply Tsallis nonextensive statistical mechanics to investigate CMB spectrum and related cosmological processes. Based on recent observational data we find that the nonextensive statistical mechanics can modify the values of related physical quantites. Since the value of physical quantites have changed, some processes, such as recombination, can be affected. We have investigated the anisotropy of the CMB for two effects: the dipole anisotropy of CMB and the Sunyaev-Zel’dovich effect. We find that the dipole anisotropy of CMB cannot be modified by the nonextensive statistical mechanics. However, the standard result of the Sunyaev-Zel’dovich effect should be modified by nonextensive statistical mechanics. In principle, future work can distinguish these effects.}

\begin{document} 
\maketitle
\flushbottom

\section{Introduction}
\label{sec:intro}

Before mid-1960s, the most significant part of the information on the universe came from observations of the distances of distant galaxies and redshifts [1]. In 1965, a nearly isotropic background of microwave radiation, i.e. Cosmic Microwave Background (CMB), was discovered by Robert Wilson and Arno Penzias, which has provided a wealth of new cosmological data [1,2]. The cosmic microwave background explains well the radiation relic from the development of the early universe, and its discovery is considered a milestone in the Big Bang universe model . Any proposed model of the universe must account for this radiation, thus the cosmic microwave background has become the key to accurate measurement of cosmology [1,2].\\
CMB radiation is almost isotropic, but the subtle fluctuations show anisotropy. This is a very active research area, and researchers are simultaneously seeking better data (for example, Planck satellites) and better initial conditions for the expansion of the universe [3,4,5]. Let us explain the mechanism of generating CMB. The work done by pressure in an expanding fluid uses heat energy drawn from the fluid [1,2]. Since the universe is expanding, we can expect that in the past time matter was hotter and denser than at present. At sufficiently early times, the rapid collisions of photons with free electrons would have kept radiation in thermal equilibrium with the hot dense matter [1]. According to the standard statistical mechanics, we can find that the photon number density in thermal equilibruim with matter at temperature $T$ between frequency interval $\nu$ and $\nu + d\nu$ is given by the blackbody radiation spectrum, namely [1,2],
\begin{equation}\label{eq:1.1}
n_T(\nu) d\nu = \frac{8\pi \nu^2 d\nu}{\exp (h \nu /k_B T) -1},
\end{equation}
where $h$ is the Planck's constant and $k_B$ is the Boltzmann constant.\\
Then the matter became less dense and cooler as time passed, and the radiation began a free expansion eventually, while its spectrum has still kept the same form [1,2]. This can be understood under an assumption that there was a $t_L$ when radiation suddenly went from being in thermal equilibrium with matter to a free expansion. Under this assumption, we can obtain the photon number density at time $t$ eventually [1]:
\begin{equation}\label{eq:1.2}
n(\nu,t) d\nu = \frac{8\pi \nu^2 d\nu}{\exp (h \nu /k_B T(t)) -1}=n_{T(t)} (\nu) d\nu,
\end{equation}  
where
\begin{equation}\label{eq:1.3}
T(t) =T(t_L) a(t_L)/ a(t).
\end{equation}
Therefore, the photon number density has been given by the blackbody radiation form even if the photons went out of equilibrium with matter, but with a redshifted temperature $(1.3)$ [1].\\
Then let us discuss nonextensive statistical mechanics briefly. One of Gibbs' arguments is that the Boltzmann-Gibbs (BG) theory cannot be applied in systems with divergency in the partition function, such as the gravitational system [6]. Therefore, the standard statistical mechanics is not universal and can only be applied for extensive systems. A generalized version of statistical mechanics has been proposed by Tsallis which can be used to explain the properties of system involving gravitational effect [7,8,9,10]. So far, the so-called “nonextensive” statistical mechanics has been applied to investigate various physical systems, such as two-dimensional turbulence, linear response theory, solar neutrino problem [11,12,13] etc. In particular, nonextensive statistical mechanics has been applied to study the properties of black holes and the universe in recent years [7,8,9,10].\\
The generalized form of entropy is [14] 
\begin{equation}\label{eq:1.4}
S_q = k_B \frac{1- \sum_{i=1}^{W} p^q_i}{q-1}, \qquad q \in \textbf{R},
\end{equation}
where $k_B$ is Boltzmann constant, $W$ is the total number of physical states of the system, and the set of probabilities $p_i$ satisfies
\begin{equation}\label{eq:1.5}
\sum_{i=1}^{W} p_i =1.
\end{equation}
Eq.(1.4) can be recovered to Boltzmann-Maxwell form if we take the limit $q \rightarrow 1$, i.e. [14],
\begin{equation}\label{eq:1.6}
\lim_{q \rightarrow 1} S_q = -k_B \sum_{i=1}^{W} p_i \ln p_i.
\end{equation}
Due to the significance of CMB for cosmology, in this article we intend to investigate the properties of CMB further. As we have pointed out before, the Boltzmann-Gibbs (BG) theory cannot be applied in systems with divergency in the partition function, such as the gravitational system [6], we expect that the blackbody radiation spectrum should be modified considering the effect of nonextensive statistical mechanics. Thus, the standard results for CMB of the universe should be modified as well. However, the effects have not been discussed yet. Therefore, in this article, we will study modifications of CMB spectrum by nonextensive statistical mechanics. The article is composed as follows: in section 2, we briefly review nonextensive thermostatistical investigation of the blackbody radiation. In section 3, we calculate the expectations of CMB. In section 4, we briefly consider the effect of recombination. In section 5, we check if nonextensive statistical mechanics can influence the dipole anisotropy. In section 6, the Sunyaev–Zel’dovich effect has been considered. The results we have obtained are discussed in section 7.\\ 

\section{Review of nonextensive thermostatistical investigation of the blackbody radiation}
Now let us briefly review the basic result of the blackbody radiation considering the effect of nonextenstive statistical mechanics. Cosmic Microwave Background is the relic of the early universe, which can be regarded as blackbody radiation. While blackbody radiation is thought of photon gas which is in thermal equilibruim and without interaction. According to ref.[15], in nonextensive statistical mechanics, the most probable distribution over a single state is given by:
\begin{equation}\label{eq:2.1}
\bar{n}(g,q) = \frac{1}{[1-(1-q)\beta (\epsilon_k - \mu)]^{1/(q-1)} +2g-1},
\end{equation}
where $g$ is a statistic number, $\beta = 1/(k_B T)$, $k_B$ is Boltzmann constant, $\mu$ is the chemical potential. When $q \rightarrow 1$, Bose-Einstein distribution for $g=0$ and Fermi-Dirac distribution for $g=1$ can be recovered [15].\\
Since the photon number is not conservative, we have $\mu=0$. Photon is boson, therefore $g=0$. Considering the energy of photon is given by $\epsilon_k = \hbar \omega_k$, where $\omega_k$ is the frequency of photon, then the most probable distribution of photon gas should be rewriten as:
\begin{equation}\label{eq:2.2}
\bar{n}(0,q) = \frac{1}{[1-(1-q)\beta \hbar \omega ]^{1/(q-1)} -1}.
\end{equation}  
For a sufficiently large volume, discrete distribution can be treated as continuous distribution of eigenfrequencies of radiation. The density of states of the photons between $\omega$ and $\omega + d\omega$ in a volume $V$ is given by [15]:
\begin{equation}\label{eq:2.3}
D(\omega) = \frac{V \omega^2}{\pi^2 c^3}.
\end{equation}
Then number of photons in the frequency interval $d\omega$ can be obtained by $dN_q(\omega) = \bar{n}(0,q) D(\omega) d\omega$, namely,
\begin{equation}\label{eq:2.4}
dN_q(\omega) = \frac{V}{\pi^2 c^3} \frac{\omega^2 d\omega}{[1-(1-q)\beta \hbar \omega ]^{1/(q-1)} -1}.
\end{equation}
Then the radiation energy interval $dU_q(\omega) = \hbar \omega dN_q(\omega)$ is :
\begin{equation}\label{eq:2.5}
dU_q(\omega) = \frac{V \hbar}{\pi^2 c^3} \frac{\omega^3 d\omega}{[1-(1-q)\beta \hbar \omega ]^{1/(q-1)} -1}.
\end{equation}
Now let us calculate the thermodynamic quantities of blackbody radiation in nonextensive statistical mechanics, such as free energy, entropy, specific heat, pressure etc. \\
The free energy $F_q$ for chemical potential $\mu = 0$ is given by [15]:
\begin{equation}\label{eq:2.6}
F_q = -k_B T \sum_{i=1}^{\infty} \ln \{\sum_{n=0}^{\infty} [1-(1-q)n \beta \hbar \omega_i]^{1/(1-q)}\},
\end{equation}
namely,
\begin{equation}\label{eq:2.7}
F_q = -k_B T \sum_{i=1}^{\infty} \ln \frac{1}{[1-(1-q) \beta \hbar \omega_i]^{1/(1-q)} }.
\end{equation}
Considering the density of states, i.e., eq.(2.3), we have [15]
\begin{equation}\label{eq:2.8}
F_q = \frac{V}{\pi^2 c^3} \int d\omega \omega^2 {[1-(1-q) \beta \hbar \omega]^{1/(1-q)}}.
\end{equation}
We can define a new variable of integration $x= \hbar \omega /k_B T$, then we have:
\begin{equation}\label{eq:2.9}
F_q = -V \frac{(k_B T)^4}{3 \pi^2 \hbar^3 c^3} I_q(3),
\end{equation}
where
\begin{equation}\label{eq:2.10}
I_q(3) = \int_{0}^{\infty} \frac{x^3}{{[1-(1-q)x]^{1/(1-q)} -1}} dx,
\end{equation}
for both $q>1$ and $q<1$ and [15] 
\begin{equation}\label{eq:2.11}
I_q(3) = \frac{\pi^4}{15} \frac{1}{(4-3q)(3-2q)(2-q)}.
\end{equation}
In order to avoid singularities, the largest value of $q$ is $\frac{4}{3}$.\\ 
Inserting (2.11) into (2.9) we obtain that [15]:
\begin{equation}\label{eq:2.12}
F_q = - \frac{4V}{3c} a_q T^4,
\end{equation}
where
\begin{equation}\label{eq:2.13}
a_q = \frac{a_B}{(4-3q)(3-2q)(2-q)}
\end{equation}
and 
\begin{equation}\label{eq:2.14}
a_B = \frac{\pi^2 k^4}{60 \hbar^3 c^2}
\end{equation}
is the so-called Stefan-Boltzmann constant. From eq.(2.13) we can find that $a_q$ is independent of temperature.\\
If we do Taylor expansion for (2.13) around $q=1$, we obtain that [15]
\begin{equation}\label{eq:2.15}
a_q = a_B [1 + 6(q-1) + 25(q-1)^2 + O(q-1)^3].
\end{equation}
When $q \rightarrow 1$, the standard Stefan-Boltzmann constant $a_B$ can be recovered.\\
Then the entropy of blackbody radiation is [15]
\begin{equation}\label{eq:2.16}
S_q = - \left(\frac{\partial F_q}{\partial T}\right)_V = \frac{16V}{3c} a_q T^3.
\end{equation}
Therefore, the total blackbody radiation energy is given by [15]
\begin{equation}\label{eq:2.17}
U_q = F_q + TS_q = -3F_q = \frac{4V}{c} a_q T^4 = V \frac{(k_B T)^4}{\pi^2 \hbar^3 c^3} I_q(3).
\end{equation}
According to eq.(2.17), we can obtain the specific heat of blackbody radiation:
\begin{equation}\label{eq:2.18}
C^q_V = \left(\frac{\partial U_q}{\partial T}\right)_V = \frac{16V}{c} a_q T^3.
\end{equation}
Moreover, we can obtain pressure from eq.(2.12), namely,
\begin{equation}\label{eq:2.19}
P_q = -\left(\frac{\partial F_q}{\partial V}\right)_T = \frac{4}{3c} a_q T^4.
\end{equation}
Then we have 
\begin{equation}\label{eq:2.20}
P_q V= \frac{1}{3} U_q,
\end{equation}
which is the equation of state for photon gas in nonextensive statistical mechanics [15].\\
By doing integration of eq.(2.4), we can obtain the mean total number of photons in a blackbody radiation, namely [15],
\begin{equation}\label{eq:2.21}
N_q =\int_{0}^{\infty} dN_q(\omega) = \frac{V}{\pi^2 c^3} \int_{0}^{\infty} \frac{\omega^2 d\omega}{[1-(1-q)\beta \hbar \omega ]^{1/(q-1)} -1}.
\end{equation}
If we define a new variable $x= \hbar \omega /k_B T$, then we have 
\begin{equation}\label{eq:2.22}
N_q = \frac{VT^3}{\pi^2 c^3 \hbar^3} I_q(2),
\end{equation}
where
\begin{equation}\label{eq:2.23}
I_q(2)= \int_{0}^{\infty} \frac{x^2}{[1-(1-q)x]^{1/(q-1)} -1}.
\end{equation}
$I_q(2)$ has an explicit expression, 
\begin{equation}\label{eq:2.24}
I_q(2)= \frac{I}{(3-2q)(2-q)}, \qquad for \quad q<1 \quad and \quad q>1,
\end{equation}
where 
\begin{equation}\label{eq:2.25}
I= \int_{0}^{\infty}\frac{x^2}{e^x -1} dx =\Gamma(3) \zeta(3).
\end{equation}
Inserting eq.(2.23) into eq.(2.22), one can find [15]
\begin{equation}\label{eq:2.26}
N_q = \frac{N}{(3-2q)(2-q)},
\end{equation}
where
\begin{equation}\label{eq:2.27}
N= \frac{V}{\pi^2} \left(\frac{T}{\hbar c} \right)^3 \Gamma(3) \zeta(3).
\end{equation}
is the standard result of photon number.\\
If we do Taylor expansion for eq.(2.26) around $q=1$, we have
\begin{equation}\label{eq:2.28}
N_q = N [1 + 3(q-1) + 7(q-1)^2 + O(q-1)^3].
\end{equation}
When $q \rightarrow 1$, the standard photon number $N$ can be recovered, i.e., eq.(2.27).\\ 

\section{Expectations of Cosmic Microwave Background}
In this section, we will calculate some basic expectations of Cosmic Microwave Background considering nonextensive statistical mechanics at first. Since the standard statistical mechanics can describe our universe accurately, we can conclude that $|q-1|$ must be close to 0. According to the results in section 2, we know that the energy density in CMB is given by
\begin{equation}\label{eq:3.1}
\int^{\infty}_{0} h\nu n(\nu) d\nu = a_q T^4 \approx a_B [1 + 6(q-1)] T^4,
\end{equation}
where we only consider $(q-1)$ order and neglect higher orders. In c.g.s units, the Stefan-Boltzmann constant $a_B$ is [1]
\begin{equation}\label{eq:3.2}
a_B = \frac{8\pi^5 k^4_B}{15 h^3 c^3} = 7.56577(5) \times 10^{-15} erg \cdot cm^{-3} \cdot deg^{-4}.
\end{equation}
Therefore, considering the effect of nonextensive statistical mechanics, the energy density in CMB is
\begin{equation}\label{eq:3.3}
\int^{\infty}_{0} h\nu n(\nu) d\nu = 7.56577(5) \times 10^{-15} [1 + 6(q-1)] erg \cdot cm^{-3} \cdot deg^{-4}.
\end{equation}
Based on observational data, we can obtain the temperature $T$ of CMB is about $2.725K$ [1]. Inserting $T_{\gamma 0}=2.725K$ into eq.$(3.3)$, an equivalent mass density (reverting to c=1) is given by
\begin{equation}\label{eq:3.4}
\rho_{\gamma 0} = a_B T^4_{\gamma 0} = 4.64 \times 10^{-34} [1 + 6(q-1)] g \cdot cm^{-3}. 
\end{equation}
For any value of the Hubble constant $H_0 \equiv \dot{a}(t_0)/a(t_0)$, we may define a critical present density [1,2,3,4,5]
\begin{equation}\label{eq:3.5}
\rho_{0, crit} \equiv \frac{3H^2_0}{8\pi G} =1.878 \times 10^{-29} h^2 g/cm^3,
\end{equation}
where $h$ is the Hubble constant in units of $100 km \cdot s^{-1} \cdot Mpc^{-1}$. \\
Taking the ratio of $\rho_{\gamma 0}$ with the critical density $\rho_{0, crit}$, we have
\begin{equation}\label{eq:3.6}
\Omega_{\gamma} \equiv \frac{\rho_{\gamma 0}}{\rho_{0, crit}} = 2.47 \times 10^{-5} [1 + 6(q-1)] h^{-2}.
\end{equation}
According to the results of cosmology, we know that photons are accompanied with neutrinos and antineutrinos of three generations, hence the total energy density in CMB is [1]
\begin{equation}\label{eq:3.7}
\rho_{R0} = [1 + 3 \times \frac{7}{8} \times (\frac{4}{11})^{4/3}] \rho_{\gamma 0} = 7.80 \times 10^{-34} [1 + 6(q-1)] g/cm^{3}.
\end{equation}
In other words, using critical density $\rho_{0, crit}$, we have
\begin{equation}\label{eq:3.8}
\Omega_{R} \equiv \frac{\rho_{R 0}}{\rho_{0, crit}} = 4.15 \times 10^{-5} [1 + 6(q-1)] h^{-2}.
\end{equation}
We can obtain the photon number density of CMB as well. According to eq.$(2.4)$ and $(2.28)$, we have
\begin{equation}\label{eq:3.9}
n_{\gamma 0} = \frac{30 \zeta(3)}{\pi^4} \frac{a_q T^3}{k_B} = 0.3702 [1 + 3(q-1)] \frac{a_B T^3}{k_B} = 20.28 [1 + 3(q-1)] [T(deg K)]^3 cm^{-3},
\end{equation}
where $\zeta(3) = 1.202057$ and we have neglected higher orders [15]. For $T=2.725K$ this gives the photon number
\begin{equation}\label{eq:3.10}
n_{\gamma 0} = 410 [1 + 3(q-1)] \quad photons/cm^3. 
\end{equation}
We can find that the above basic physical quantities in CMB have been modified by order $(q-1)$. \\
There is an effect of the cosmic microwave background that has long been expected but has been difficult to observe. At moderate energy, a cosmic ray proton striking a photon in the CMB can only scatter the photon. The rate of the process is proportional to $\alpha^2$, where $\alpha \approx 1/137$ is the fine structure constant. However, at high energy, in the reactions, such as $\gamma + p \rightarrow \pi^0 + p$ or $\gamma + p \rightarrow \pi^{+} + n$, processes whose rate is proportional to $\alpha$ [1]. If we assume that the high energy cosmic radiation come from outside the milkway galaxy, hence we expect a dip in the spectrum of cosmic ray protons at an energy where the cross section for these processes becomes obvious [1]. The center of mass energy is given by
\begin{equation}\label{eq:3.11}
W = \left( (q + \sqrt{p^2 + m^2_p} )^2 - |\textbf{q}+ \textbf{p}|^2 \right)^{1/2} \approx (2qp(1-\cos\theta) + m^2_p)^{1/2}, 
\end{equation}
where $\textbf{q}$ and $\textbf{p}$ are the initial photon and proton momenta ($p \gg m_p \gg q$) and $\theta$ is the angle between these two momenta [1]. The threshold $W > m_{\Delta}$, where $m_{\Delta} = 1232 MeV$ is the mass of the pion–nucleon resonance with isospin $3/2$ and spin-parity $3/2^{+}$, requires that 
\begin{equation}\label{eq:3.12}
2qp(1-\cos\theta) > m^2_{\Delta} - m^2_p.
\end{equation} 
The typical energy in CMB at temperature $T_{\gamma 0} =2.725K$ is $\rho_{\gamma 0}/ n_{\gamma 0} \approx 6 \times 10^{-4} \frac{[1+6(q-1)]}{[1+3(q-1)]}  eV$ and the largest value of $1-\cos \theta$ is 2, therefore the threshold is about
\begin{equation}\label{eq:3.13}
p_{threshold} \approx \frac{m^2_{\Delta} - m^2_p}{4 \rho_{\gamma 0}/ n_{\gamma 0} } \approx 2.658 \times 10^{20} [1+3(q-1)]/[1+6(q-1)] eV,
\end{equation}
where $[1+3(q-1)]/[1+6(q-1)]$ is the modification term. \\
The effect is difficult to see. However, some groups have announced that they have observed this effect, namely, “sharp suppression” of the primary cosmic ray spectrum [16,17]. In particular, in ref.[16,17] reported the suppression at $6 \times 10^{19}eV$. \\
According to recent observational data [18], we have that $q=1^{+0.05}_{-0.12}$. Thus, we can obtain $ -0.72<6(q-1)< 0.30$, $ -0.36<3(q-1)< 0.15$ and $0.8846 <[1+3(q-1)]/[1+6(q-1)]<2.2857 $. We can find that the effect of nonextensive statistical mechanis can modify the value of physical quantities, such as energy density, photon number density and threshold energy, significantly. \\
Moverover, at the time that the density of the universe was dominated by photons and neutrinos [1], the Hubble constant is given by  
\begin{equation}\label{eq:3.14}
H \equiv \frac{\dot{a}}{a} = H_0 \sqrt{\Omega_{R} T^4 / T^4_{\gamma 0}} =2.1 \times 10^{-20} \sqrt{1+6(q-1)} s^{-1} \left(\frac{T}{T_{\gamma 0}}^2 \right).
\end{equation}
The value of $H$ is determined by the value of $q$. Hence, the duration of this era will change as well. \\

\section{The effect of nonextensive statistical mechanics on the recombination process}
Based on previous research, photons stopped exchanging energy effectively with electrons when the temperature of the expanding universe dropped to about $10^5K$ [1]. After that, photons continued to be scattered by free electrons. This process terminated when the free electrons became bound into hydrogen and helium atoms, ending the scattering of photons. This process is called recombination [1]. Now we briefly comment on how nonextensive statistical mechanics affects recombination process. \\
After the nucleosynthesis took place in the first few minutes, the universe continued to cool down, while the nuclei and electrons is still an ordinary plasma in thermal equilibrium with the photons. After several hundred thousand years, temperature dropped below the atomic ionization energies of a few electron volts needed for nuclei to capture electrons and form atoms. Then the universe became transparent due to the last scattering of radiation from free electrons. Thus the CMB photons became free suddenly to propagate since that time [1]. We can consider the recombination process
\begin{equation}\label{eq:4.1}
p+e \rightarrow H+\gamma.
\end{equation}  
Due to the chemical equilibrium relation, we have
\begin{equation}\label{eq:4.2}
\mu_p + \mu_e = \mu_H,
\end{equation}
where the chemical potential of blackbody photons is still zero. Since the temperature was a few thousand degrees Kelvin and there were only about $10^9$ atoms per cubic meter, the protons, electrons and hydrogen atoms can be considered classical ideal gas. Based on the standard statistical mechanics, the chemical potential of these particles are given by [19]
\begin{equation}\label{eq:4.3}
\mu_p = m_p + k_B T \ln (n_p \lambda^3_p) - k_B T \ln 2,
\end{equation}
\begin{equation}\label{eq:4.4}
\mu_e = m_e + k_B T \ln (n_e \lambda^3_e) - k_B T \ln 2,
\end{equation}
and 
\begin{equation}\label{eq:4.5}
\mu_H = m_H + k_B T \ln (n_H \lambda^3_H) - k_B T \ln 4,
\end{equation}
where $\lambda_i$ is the de Broglie wavelength.\\
Considering the nonextensive statistical mechanics, for non-relativistic particles, $m \gg T$, the number density is given by [20]
\begin{equation}\label{eq:4.6}
n_q = g \left(\frac{mT}{2\pi}\right)^{3/2} e^{-(m-\mu)/T} \times [1 + \frac{q-1}{2} \left(\frac{15}{4} + 3\frac{m-\mu}{T} + \frac{(m-\mu)^2}{T^2} \right)],
\end{equation}
where $g$ is the number of one particle's spin states, $m$ is the particle mass and $\mu$ is the chemical potential of the particle and we take $k_B =1$. \\
After immediate calculation, we can obtain the expression for $n^H_q/n^e_q n^p_q$ [20]:
\begin{equation}\label{eq:4.7}
\frac{n^H_q}{n^e_q n^p_q} = \frac{g_H}{g_e g_p} \left( \frac{m_e T}{2\pi} \right)^{-3/2} \exp \{(m_p + m_e- m_H)/T \} \times [ \frac{ \bar{u} ( (m_H-\mu_H) /T) }{ \bar{u} ( (m_p-\mu_p) /T) \bar{u} ( (m_e-\mu_e) /T) } ],
\end{equation}
where we have defined
\begin{equation}\label{eq:4.8}
\bar{u}_i \equiv \bar{u} ( (m_p-\mu_p) /T) \equiv 1 + \frac{q-1}{2} \left(\frac{15}{4} + 3\frac{m_i-\mu_i}{T} + \frac{(m_i-\mu_i)^2}{T^2} \right).
\end{equation}
We should point out that the chemical potential $\mu_i$ in the above formula are the standard chemical potential, i.e., $\mu_i$ is obtained from the standard statistical mechanics. We can write the generalized chemical potential as $\mu_q = \mu_{st} + (q-1)C$, where $\mu_q$ is the generalized chemical potential while $\mu_{st}$ is the standard chemical potential and $C$ is a factor including second order correction [20]. If we only consider the lowest order correction, the standard chemical potential can still be used in these formula.\\
Due to charge neutrality, the number density of electrons and protons should be the same, namely [1,2,20],
\begin{equation}\label{eq:4.10}
n_e = n_p.
\end{equation}
Using $g_p=g_e =2$, $g_H =4$, we can rewrite eq.$(4.7)$ as [20]
\begin{equation}\label{eq:4.11}
\frac{n^H_q}{n^B_q} = \frac{n^p_q}{n^B_q} \frac{n^e_q}{n^B_q} n^q_{\gamma} \eta_q \left( \frac{m_e T}{2\pi} \right)^{-3/2} \exp \{(m_p + m_e- m_H)/T \} \times [ \frac{ \bar{u} ( (m_H-\mu_H) /T) }{ \bar{u} ( (m_p-\mu_p) /T) \bar{u} ( (m_e-\mu_e) /T) } ],
\end{equation}
where $\eta_q= \frac{n^B_q}{n^{\gamma}_q} $, $n^q_{\gamma} = n_{\gamma, st} + (q-1) n_{\gamma,c}$ and $n_B$ is the baryon number.\\
If we define $X^e_q =\frac{n^p_q}{n^B_q} $ and $1- X^e_q =\frac{n^H_q}{n^B_q} $, then one can derive the generalized Saha law [20]:
\begin{equation}\label{eq:4.12}
\begin{aligned}
\frac{1- X^e_q}{(X^e_q)^2} 
&=\frac{1- X^e_{st}}{(X^e_{st})^2}+ \frac{q-1}{2} \frac{4\sqrt{2}}{\sqrt{\pi}} \zeta(3) \eta_{st} \left( \frac{ T}{m_e} \right)^{3/2} \\
& \times e^{B/T} \left(\frac{2 e^{(\mu_H-m_H)/T} u_H + e^{(\mu_p-m_p)/T} u_p}{2 e^{(\mu_H-m_H)/T} + e^{(\mu_p-m_p)/T} } + u_H -u_e -u_p \right), \\
\end{aligned}
\end{equation}
where $B=m_p+m_e-m_H$ is the Rydberg energy. More details can be found in ref.[20].\\
The number density of photon has been modified considering the effect of nonextensive statistical mechanics, which has been derived in section 3. Moreover, the number density of non-relativistic particles has been modified by nonextensive statistical mechanics, namely, eq.$(4.6)$. Thus, we conclude that the nonextensive statistical mechanics will affect the results of recombination process of the universe. The deviation from standard results should be determined by future observations.\\

\section{On the dipole anisotropy of CMB}
In the previous two sections, we treated CMB as perfectly isotropic and homogeneous. This is a very good approximation. However, there still exist small fluctuations and variations in different directions. The tiny variations from perfect isotropy can provide some of the most significant information about the evolution of the universe [1]. In section 5 and 6, we will consider if nonextensive statistical mechanics can make contributions to the anisotropy of cosmic microwave background. In this section, we consider the dipole anisotropy of CMB, which is the simplest and earliest detected departure from isotropy of the observed cosmic microwave background, arising from the earth’s motion [1]. In section 6 we will consider the Sunyaev–Zel’dovich effect.\\
To analyze the measurement of our own galaxy through the cosmic microwave background, it is convenient to consider the density of photons $N_{\gamma} (\textbf{p})$ in phase space, defined by specifying that there are $N_{\gamma} (\textbf{p})d^3p$ photons of each polarization (left or right circularly polarized) per unit spatial volume in a momentum-space volume $d^3 p$ centered at $\textbf{p}$ [1,2]. \\
Since the density of states between $\nu$ and $\nu + d\nu$ is $4\pi h^3 \nu^2 d\nu/c^3$ and the photon momentum $|\textbf{p}| = h\nu/c$, accroding to eq.$(2.2)$, we have
\begin{equation}\label{eq:5.1}
N_{\gamma} (\textbf{p})= \frac{1}{2} \frac{n_T(c|\textbf{p}|/h)}{4\pi h^3 \nu^2 d\nu/c^3} = \frac{1}{h^3} \frac{1}{[1-(1-q)|\textbf{p}|c/k_BT]^{1/(q-1)}-1},
\end{equation} 
where the factor $\frac{1}{2}$ is due to the fact that $n_T(c|\textbf{p}|/h)$ consists of two photon polarization states. $N_{\gamma} (\textbf{p})$ is the density which is measured by rest observer in the CMB [1]. $N_{\gamma} (\textbf{p})$ is a Lorentz scalar, namely, under a Lorentz transformation to a coordinate system moving with respect to the radiation background which takes $\textbf{p}$ to $\textbf{p}'$, we always have 
\begin{equation}\label{eq:5.2}
N'_{\gamma} (\textbf{p}')= N_{\gamma} (\textbf{p}). 
\end{equation}
The next step is write down the Lorentz transformations relating primed and unprimed quantities. Recall the Lorentz transformation relations, if the earth is moving in the three-direction, we should have [1,2]
\begin{equation}\label{eq:5.3}
\left(
\begin{array}{cccc} 
p_1\\
p_2\\
p_3\\
|\textbf{p}|
\end{array}
\right) 
=
\left(
\begin{array}{cccc} 
1 & 0 & 0 & 0\\
0 & 1 & 0 & 0\\
0 & 0 & \gamma & \beta \gamma\\
0 & 0 & \beta\gamma & \gamma
\end{array}
\right)
\left(
\begin{array}{cccc} 
p'_1\\
p'_2\\
p'_3\\
|\textbf{p}'|
\end{array}
\right),
\end{equation}
where $\beta$ is the earth's moving velocity, $\gamma \equiv (1- \beta^2)^{-1/2}$, $\textbf{p}$ is the photon momentum in a rest frame in the CMB, while $\textbf{p}'$ is the photon momentum measured on the earth [1,2]. In particular, we have
\begin{equation}\label{eq:5.4}
|\textbf{p}| = \gamma (1+ \beta \cos \theta )|\textbf{p}' |, 
\end{equation}  
where $\theta$ is the angle between the three-axis and $\textbf{p}'$. \\
Combining eqs.$(5.2)$ and $(5.4)$, we can obtain that
\begin{equation}\label{eq:5.5}
N'_{\gamma} (\textbf{p}') =  \frac{1}{h^3} \frac{1}{[1-(1-q)|\textbf{p}'|c/k_BT']^{1/(q-1)}-1}, 
\end{equation} 
where 
\begin{equation}\label{eq:5.6}
T' = \frac{T}{\gamma (1+ \beta \cos \theta )}. 
\end{equation} 
We can find that the temperature $T'$ is a function of the earth's velocity and the angle $\theta$. \\
Comparing the standard result of $T'$ [1,2,3,4,5], we can find that the two $T's$ have the same expression. Therefore, we cannot detect the effect of nonextensive statistical mechanics by the dipole anisotropy of CMB. Therefore, in the next section we will investigate another contribution to the anisotropy of CMB, i.e., the Sunyaev–Zel’dovich effect.  

\section{The Sunyaev–Zel’dovich effect and Kompaneets equation}
The Sunyaev–Zel’dovich effect, which is another contribution to the anisotropy of the CMB, is originated from the scattering of this radiation by electrons in intergalactic space within clusters of galaxies along the line of sight [1,2,21,22,23]. The equation describing the effect on the scattering of CMB by isotropic nonrelativistic electron motions was derived by Kompaneets [24], while its application in cosmology is called the Sunyaev–Zel’dovich effect, after their poineering analyses [21,22,23].\\
The Kompaneets equation shows that scattering of the CMB by a non-relativistic electron gas changes the observed photon occupation number $N(\omega)$ at photon energy $\hbar \omega \ll m_e c^2$ at a rate (in cgs units), where we have defined that $4\pi \omega^2 N(\omega) d\omega$ is the number of photons of each of the two polarization states between energy $\hbar \omega$ and $\hbar (\omega + d\omega)$, namely [1],
\begin{equation}\label{eq:6.1}
\dot{N} (\omega) =  \frac{n_e \sigma_T}{m_e c \omega^2} \frac{\partial}{\partial \omega} [ k_B T_e \omega^4 \frac{\partial N(\omega)}{\partial \omega} + \hbar \omega^4 N(\omega) (1+ N(\omega))],
\end{equation}
where $n_e$ is the number density of electrons, $T_e$ is the temperature of electrons and $\sigma_T$ is cross section of Thompson scattering. The details of derivation of the Kompaneets equation will be given in Appendix.\\
However, we are interested in the change of the appearance to us of the CMB, which is scattered by a cloud of electrons along the line of sight, therefore, the Kompaneets equation $(6.1)$ could be rewritten as [1]
\begin{equation}\label{eq:6.2}
\frac{\partial }{\partial l} N(\omega,l) =  \frac{n_e (l) \sigma_T}{m_e c^2 \omega^2} \frac{\partial}{\partial \omega} [ k_B T_e (l) \omega^4 \frac{\partial N(\omega,l)}{\partial \omega} + \hbar \omega^4 N(\omega,l) (1+ N(\omega,l))],
\end{equation} 
where $l$ is the proper distance coordinate along the line of sight. \\
Since the typical value for the photon energy $\hbar \omega$ is $10^{-4}eV$ to $10^{-3}eV$, while the temperature of the ionized plasma in clusters of galaxies is typically higher than $10^6$ degrees, the second term in eq.$(6.2)$ can be neglected, in other words, eq.(6.2) can be simplified to 
\begin{equation}\label{eq:6.3}
\frac{\partial }{\partial l} N(\omega,l) =  \frac{n_e (l) \sigma_T k_B T_e (l)}{m_e c^2 \omega^2} \frac{\partial}{\partial \omega} [  \omega^4 \frac{\partial N(\omega,l)}{\partial \omega} ].
\end{equation}
From eq.(6.3), we can obtain that the change in the occupation number $\Delta N(\omega)$ through the cloud is given by [1,2]
\begin{equation}\label{eq:6.4}
\Delta N(\omega) = \frac{y}{\omega^2} [  \omega^4 \frac{\partial N(\omega)}{\partial \omega} ],
\end{equation} 
where we have defined 
\begin{equation}\label{eq:6.5}
y \equiv \frac{\sigma_T}{m_e c^2} \int dl n_e(l) k_B T_e(l).
\end{equation}
Here the integral path is taken along the line of sight through the cloud. \\
Considering the effect of Tsallis nonextensive statistical mechanics, for blackbody radiation at temperature $T$, we should have 
\begin{equation}\label{eq:6.6}
N (\omega) = \bar{n}(0,q) = \frac{1}{[1-(1-q)\beta \hbar \omega ]^{1/(q-1)} -1}.
\end{equation}
According to Appendix, when we consider the effect of Tsallis nonextensive statistical mechanics, eqs.$(6.3)$ and $(6.4)$ can still be applied. Then using eq.$(6.4)$, we can obtain that
\begin{equation}\label{eq:6.7}
\begin{aligned}
\Delta N (\omega) = & y \{ \frac{-4x}{([1-(1-q)x]^{1/(q-1)}-1)^2} [1-(1-q)x]^{(2-q)/(q-1)} \\
& + \frac{2x^2}{([1-(1-q)x]^{1/(q-1)}-1)^3} [1-(1-q)x]^{(4-2q)/(q-1)} \\
& - \frac{(2-q)x^2}{([1-(1-q)x]^{1/(q-1)}-1)^2} [1-(1-q)x]^{(3-2q)/(q-1)} \},
\end{aligned}
\end{equation}
where we have defined $x \equiv \hbar \omega/ k_B T$. In principle, astronomers can use eq.(6.7) to distinguish the effect between the anisotropies of CMB due to the primary anisotropies and the Sunyaev–Zel’dovich effect [1,2]. \\
The change in the photon occupation number $\Delta N (\omega)$ obtained from the standard statistical mechanics is given by [1,2]
\begin{equation}\label{eq:6.8}
\Delta N (\omega) = y \{ \frac{-x + (x^2/4) \coth (x/2)}{\sinh^2 (x/2)} \},
\end{equation} 
where $x=\hbar \omega/ k_B T$. In principle, people can investigate the effect of nonextensive statistical mechanics by comparing the results of eq.$(6.7)$ and $(6.8)$. In the Rayleigh-Jeans part of the blackbody spectrum, namely, $x \ll 1$, both $(6.7)$ and $(6.8)$ give the same limit, i.e., $\Delta N \rightarrow -2y/x$. 

\section{Conclusions and outlook}
The cosmic microwave backround (CMB) can supply us significant information of the universe. The radiation is almost isotropic, but the subtle fluctuations show anisotropy. Most cosmologists believe that the Big Bang model explains the cosmic microwave background best
[1,2,3,4,5]. Therefore, one important step in cosmology is to investigate the property of CMB further. \\
Moreover, many researchers have belived that the standard statistical mechanics cannot be applied in nonextensive systems, such as gravitational system [6]. A generalized statistical mechanics has been proposed by Tsallis to study nonextensive systems [7-14]. Since the universe is a gravitational system, we expect that the standard results in cosmology should be modified considering the effect of nonextensive statistical mechanics. Due to the significance of CMB, in this article we have investigated how the CMB spectrum should be modified in Tsallis statistical mechanics. \\
In this article, we have considered four sections. In section 3, we have obtained the expectations of Cosmic Microwave Background for basic physical quantities. We can find that the deviation from the standard statistical mechanics can change the value of these physical quantities significantly. Based on the results of section 3, the change of photon number density will affect the details of cosmological processes. In section 4, we briefly consider the recombination process including the effect of nonextensive statistical mechanics. In section 3 and 4, we treat CMB as perfectly isotropic and homogeneous. This is a very good approximation. However, there still exist small fluctuations and variations in different directions. The tiny variations from perfect isotropy can provide us some of the most significant information about the evolution of the universe [1,2,3,4,5]. Therefore, in section 5 and 6, we investigat two anisotropic effects of CMB: the dipole anisotropy of CMB and the Sunyaev-Zel'dovich effect. In section 5, we have found that nonextensive statistical mechanics does not influence the result of the dipole anisotropy of CMB. In section 6, we find that the difference of the photon occupation number $N(\omega)$ between two versions of statistical mechanics. In principle, one can investigate the effect of nonextensive statistical mechanics by comparing the results in the standard statistical mechanics and Tsallis statistical mechanics. \\
Future work can be directed along at least three lines of further research. Firstly, blackbody radiation satifies Bose-Einstein distribution, it is natural to investigate Fermi-Dirac distribution in nonextensive statistical mechanics. Secondly, we should consider Fermion systems in the universe considering the effect of nonextensive statistical mechanics. Thirdly, there is a close relationship between statistical mechanics and quantum field theory, therefore, further investigations on generalized quantum field theory should be conducted. Then we will apply generalized quantum field theory to study some physical proceses, such as Hawking radiation.

\appendix
\section{The derivation of the Kompaneets equation}
In this appendix we will briefly review the derivation of the Kompaneets equation considering the effect of nonextensive statistical mechanics [1,2]. We will find that the original Kompaneets equation can still be used to study the Sunyaev–Zel’dovich effect at order $(q-1)$.\\
Assume that an electron which has $4$-momentum
\begin{equation}\label{eq:A1}
p=(0,0,p_e, E_e), \qquad E_e \equiv \sqrt{m^2_e + p^2_e},
\end{equation} 
is travelling in the three-direction and is struck by a photon. If the initial and final energy of the photon are $\omega$ and $\omega'$, then initial $4$-momenta $q$ and final $4$-momenta $q'$ are
\begin{equation}\label{eq:A2}
q=(\sin\eta \cos\phi, \sin\eta \sin\phi, \cos\eta, 1)\omega,
\end{equation}  
\begin{equation}\label{eq:A3}
q=(\sin\eta' \cos\phi', \sin\eta' \sin\phi', \cos\eta', 1)\omega',
\end{equation}
where $\eta$ and $\phi$ are the polar and azimuthal angles for initial energy $\omega$, while $\eta'$ and $\phi'$ are the polar and azimuthal angles for final energy $\omega'$. \\
In electron rest frame, the initial and final photon four-momenta are $Lq$ and $Lq'$, where $L$ is the Lorentz transformation
\begin{equation}\label{eq:A.4}
L
=
\left(
\begin{array}{cccc} 
1 & 0 & 0 & 0\\
0 & 1 & 0 & 0\\
0 & 0 & \gamma & -\beta \gamma\\
0 & 0 & -\beta\gamma & \gamma
\end{array}
\right),
\end{equation}
where $\gamma \equiv (1-\beta^2)^{-1/2}$ and $\beta \equiv p_e/E_e$ is the electron velocity. Therefore, in the electron rest frame, the initial and final photon $4$-momenta are given by 
\begin{equation}\label{eq:A5}
Lq=(\sin\alpha \cos\phi, \sin\alpha \sin\phi, \cos\alpha, 1)k,
\end{equation} 
\begin{equation}\label{eq:A6}
Lq'=(\sin\alpha' \cos\phi', \sin\alpha' \sin\phi', \cos\alpha', 1)k',
\end{equation} 
where $k$ and $k'$ are the initial and final photon energies in the electron rest frame which are given by
\begin{equation}\label{eq:A7}
k=(-\beta \gamma \cos \eta+ \gamma )\omega,
\end{equation} 
\begin{equation}\label{eq:A8}
k'=(-\beta \gamma \cos \eta' + \gamma )\omega'.
\end{equation} 
There is no change in the azimuthal angles, while the initial and final polar angles of the photon’s velocity in this frame are
\begin{equation}\label{eq:A9}
\cos\alpha =\frac{\cos\eta - \beta}{1-\beta \cos\eta},
\end{equation} 
\begin{equation}\label{eq:A10}
\cos\alpha' =\frac{\cos\eta' - \beta}{1-\beta \cos\eta'}.
\end{equation} 
Then the fractional change in the photon energy in the original frame of reference can be given by
\begin{equation}\label{eq:A11}
\frac{\omega'-\omega}{\omega} = \frac{1}{1+ \beta \cos \alpha} \left(\frac{1+ \beta \cos \alpha'}{1+ (k/m_e) (1-\cos\theta)} -\beta \cos \alpha -1 \right).
\end{equation} 
For most cases, we have $k_e/m_e \ll 1$ and $\beta \ll 1$. Then we can obtain that
\begin{equation}\label{eq:A12}
\frac{\omega'-\omega}{\omega} \approx -(k/m_e) (1-\cos\theta) + \frac{\beta(\cos\alpha'- \cos\alpha)}{1+\beta\cos\alpha}.
\end{equation}
If photon energy $k\ll m_e$, for a rest electron, the differential cross-section is 
\begin{equation}\label{eq:A13}
d\sigma = \frac{3\sigma_T}{16\pi} (1+\cos^2 \theta) d(\cos\alpha') d\phi',
\end{equation}
where the Thomson cross section $\sigma_T$ is $e^4/ 6\pi m^2_e$. Then we can obtain the average photon energy change per collision is
\begin{equation}\label{eq:A14}
\langle \omega' - \omega \rangle =- \frac{\beta \omega \cos\alpha}{1+\beta\cos\alpha} - \frac{k\omega}{m_e}.
\end{equation}
Moreover, we have
\begin{equation}\label{eq:A15}
\langle \langle \omega' - \omega \rangle \rangle /\omega \approx \frac{4}{3} \beta^2 - \frac{k}{m_e} \approx \frac{4}{3} \beta^2 - \frac{\omega}{m_e},
\end{equation}
\begin{equation}\label{eq:A16}
\langle \langle (\omega' - \omega)^2 \rangle \rangle /\omega \approx \frac{2}{3} \beta^2 \omega^2.
\end{equation}
All these equations are for an electron at a fixed speed $\beta$ and these are results of kinectics and dynamics which are independent of statistical mechanics. If the number of electrons with speed between $\beta$ and $\beta + d\beta$ is given by the Maxwell–Boltzmann distribution with electron temperature $T_e$, and is hence proportional to $\beta^2 \exp(-m_e \beta^2 / 2 k_B T_e) d\beta$, the avarage value of $\beta^2$ is $3k_BT_e/m_e$ [1,19]. Thus, we have
\begin{equation}\label{eq:A17}
\langle \langle \omega' - \omega \rangle \rangle /\omega \approx \frac{4k_BT_e}{m_e} \omega - \frac{\omega^2}{m_e},
\end{equation}
\begin{equation}\label{eq:A18}
\langle \langle (\omega' - \omega)^2 \rangle \rangle /\omega = \frac{2k_BT_e}{m_e}\omega^2 .
\end{equation}
We define that $N(\omega)$ is the the number of photons per quantum state of energy $\omega$. Then after long calculations, we can find that the change of photon occupation number is 
\begin{equation}\label{eq:A19}
\begin{aligned}
\dot{N}(\omega) = &-\frac{n_e \sigma_T}{\omega^2} [\frac{\partial}{\partial \omega} \left( \omega^2 N(\omega) \left(1+N(\omega) \right)\langle \langle \omega' - \omega \rangle \rangle  \right) \\
& -\frac{1}{2} \frac{\partial^2}{\partial \omega^2} \left( \omega^2 N(\omega) \left(1+N(\omega) \right)\langle \langle (\omega' - \omega)^2 \rangle \rangle  \right) \\
& + \frac{\partial}{\partial \omega} \left( \omega^2 N(\omega) \langle \langle (\omega' - \omega)^2 \rangle \rangle \frac{\partial N(\omega)}{\partial \omega} \right) ].
\end{aligned}
\end{equation}
Inserting eqs.(A.17) and (A.18) into (A.19), we can obtain the Kompaneets equation:
\begin{equation}\label{eq:A20}
\dot{N} (\omega) =  \frac{n_e \sigma_T}{m_e c \omega^2} \frac{\partial}{\partial \omega} [ k_B T_e \omega^4 \frac{\partial N(\omega)}{\partial \omega} + \hbar \omega^4 N(\omega) (1+ N(\omega))].
\end{equation}
If we consider nonextensive statistical mechanics, eqs.(A.17) and (A.18) will add some corrections. If we only consider the case of $q \rightarrow 1$, corrections are small, i.e., higher order corrections can be neglected. Furthermore, $(q-1)$ order can only supply second order correction to eqs.$(6.3)$ and $(6.4)$. Therefore even if we consider the effect of nonextensive statistical mechanics, eqs.$(6.3)$ and $(6.4)$ can still be applied. 

\acknowledgments

We acknowledge beneficial discussions with Francesc Cunillera and Chad Briddon.  



\end{document}